\documentclass[12pt, draftclsnofoot, onecolumn]{IEEEtran}
\IEEEoverridecommandlockouts
\usepackage{cite}
\usepackage{amsmath,amssymb,amsfonts,bbm}

\usepackage{textcomp}
\usepackage{xcolor}
\usepackage{comment}
\def\BibTeX{{\rm B\kern-.05em{\sc i\kern-.025em b}\kern-.08em
    T\kern-.1667em\lower.7ex\hbox{E}\kern-.125emX}}
\newcommand\mydots{\hbox to 0.75em{.\hss.\hss.}}

\usepackage{amsthm}
\usepackage{ textcomp }
\usepackage{subfigure}
\usepackage{bm}
\usepackage{multirow}
\usepackage{longtable}
\usepackage{amssymb}
\usepackage{graphicx}
\usepackage{xcolor}
\usepackage{algorithm,algorithmic}
\usepackage[algo2e,ruled,vlined]{algorithm2e}
\newtheorem{prop}{Proposition}
\begin{document}
% \setlength{\abovedisplayskip}{3pt}
% \setlength{\belowdisplayskip}{3pt}
% \title{Decentralized Processing in Extra-Large MIMO Arrays: Merging Belief Propagation and Mean Field
% Approximation}
\title{Uncoordinated and Decentralized Processing in Extra-Large MIMO Arrays}
% \author{
% \IEEEauthorblockN{Abolfazl Amiri, Sajad Rezaie, Carles Navarro Manch\'on, Elisabeth de Carvalho  }\\
% \IEEEauthorblockA{ Department of Electronic Systems, Aalborg University, Denmark\\
% Email: \{aba,sre,cnm,edc\}@es.aau.dk}
% }
 \author{Abolfazl Amiri, Carles Navarro Manch\'on, Elisabeth de Carvalho 
 \thanks{The authors are with the Department of Electronic Systems, Aalborg University, 9220 Aalborg, Denmark (e-mail:
 {aba,cnm,edc}@es.aau.dk). This work was supported by the
Danish Council for Independent Research DFF-701700271.
 }
 }
\maketitle

\begin{abstract}
We propose a decentralized receiver for extra-large multiple-input multiple-output (XL-MIMO) arrays. Our method operates with no central processing unit (CPU) and all the signal detection tasks are done in distributed nodes. We exploit a combined message-passing framework to design an uncoordinated detection scheme that overcomes three major challenges in the XL-MIMO systems: computational complexity, scalability and non-stationarities in user energy distribution.  Our numerical evaluations show a significant performance improvement compared to benchmark distributed methods while operating very close to the centralized receivers.
\end{abstract}

 \begin{IEEEkeywords}
 Massive MIMO, Message-passing, decentralized receivers, XL-MIMO
 \end{IEEEkeywords}

\section{Introduction}
Beyond fifth-generation (B5G) multi-input multiple-output (MIMO) systems will rely on antenna arrays with an extreme number of elements that provide a very high spatial resolution. Recently, different variants of such systems such as   extra-large  scale  MIMO  (XL-MIMO)  systems \cite{xlmimo_mag},  or  large intelligent  surfaces  (LIS)  \cite{lis_pos}  have been introduced.
These technologies offer a boost in the system's spectral efficiency thanks to their ability to jointly serve a large number of users. However, the large array dimensions bring about three important challenges: an increase in the computational complexity of the receiver processing, difficulties in the scalability of the array architecture, and the emergence of spatial non-stationarities (NS) of the received signal energy across the array elements.

% In order to increase the dimensions of a MIMO array, we have to deal with three main challenges: computational complexity, scalability and non-stationarities (NS) on the received energy at the base station (BS). 
Most of the conventional receiver designs, e.g. zero-forcing (ZF), rely on a central processing unit (CPU) that inverts large matrices\cite{bjornson2015optimal}. Such receivers need a vast amount of computational capacity to operate and are not suitable for our XL-MIMO system dimensions. Other distributed receiver techniques such as \cite{xlmimo_GC} try to divide some of the computations between several nodes, but still, need a CPU to supervise all the data transmission steps.

Having a central node for processing all the transmitting signals requires a dedicated link between all the antenna elements and the CPU. Managing these inter-connections is costly and hinders adding more antennas next to the existing array deployment. Therefore, scaling up the array size becomes challenging. Different hierarchical processing methods are used in \cite{amiri2020distributed,rodrigues2020low,wang2019expectation} that aim to divide the processing tasks between the CPU and local units at the \textit{sub-arrays}. However, a connection between each of these sub-arrays and the CPU is still necessary and limits the scalability of the receivers. 

The last challenge is the presence of spatial NS of the channel gains imposing variable mean energy of a given user’s signal along the array  creating visibility regions (VR) \cite{xlmimo_mag}. A VR is a subset of the antennas that hold most of the user’s received energy
and limits the performance of linear receivers \cite{xlmimo_GC}. Therefore, there is a need for smarter methods that are aware of such NS and utilize it in the receiver design process.

Most of the current literature  focuses on either lowering the computational complexity or dealing with the effect of NS. Authors in \cite{sanchez2019decentralized,ZFapp2018zhang,sarajlic2019fully} propose a fully decentralized ZF approximator that works without a CPU. However, these methods have a high processing delay and their performance is highly dependent on the even distribution of users' energy on the array. The works in \cite{wang2019expectation} and \cite{decentralized2020zhang} use expectation propagation to distribute the symbol detection between a central node and few local units.
In \cite{amiri2020distributed}, we presented an NS-aware receiver that works in a hierarchical way. Yet all of these methods rely on a central node and have a relatively high processing delay.

In this paper, we propose a decentralized receiver that works without a CPU for user symbol detection. Our method leverages an approximate inference framework based on a combination of belief propagation (BP) and variational message-passing (VMP). We use distributed nodes called local processing units (LPU) that work in parallel to calculate the local symbol estimates. These nodes are only allowed to exchange information with their neighbors, making the receiver scalable and easy to deploy. Moreover, they use a local successive interference cancellation (SIC) scheme to boost the symbol detection performance.  Our numerical results show a significant improvement compared to the other decentralized techniques while obtaining almost the same performance as the centralized benchmark methods.

\textit{Notations}: Boldface lowercase and uppercase letters are representing vectors and matrices, respectively. The
cardinality of a set is denoted by $|\mathcal{I}|$; the relative complement of $i$ in a set $\mathcal{I}$ is
denoted as $\mathcal{I}\setminus i$. Superscripts
$(\cdot)^T$ and $(\cdot)^H$ show transposition and Hermitian transposition,
respectively. The probability density function (pdf)  of a multivariate
complex Gaussian distribution 
with mean $\bm{\mu}$ and covariance
matrix $\bm{\Sigma}$ and its distribution are denoted by $\text{CN}(\cdot;\bm{\mu},\bm{\Sigma})$ and $\mathcal{CN}(\bm{\mu},\bm{\Sigma})$, respectively. $\mathcal{U}(a,b)$ shows a uniform distribution in $[a,b]$. $\mathbf{I}_M$ denotes the identity matrix of size $M$ and $\bm{1}_M$ is a vector with $M$ elements all equal to $1$.
We use $f(x)\propto g(x)$ when $f(x)=c g(x)$ for some positive constant $c$.

\section{System Model}\label{sec:system model}

We assume a narrow-band MIMO system with $K$ single-antenna users and a base station (BS) with $M$ antenna elements in the uplink transmission. User symbols are denoted with the vector
$\mathbf{x}\in\mathbb{C}^K$ with entries taking values from the complex constellation set $\mathcal{M}=\{a_1,a_2,\cdots ,a_{|\mathcal{M}|}\}$. $\mathbf{H}=[\bm{h}_1,\cdots, \bm{h}_K]\in\mathbb{C}^{M\times K}$ is the channel matrix with  column vectors $\bm{h}_k$ each of which denotes user $k$'s channel. The noise at the BS is assumed to have circularly symmetric complex Gaussian distribution $\mathbf{n}\sim \mathcal{CN}(0,\sigma_n^2\mathbf{I}_M)\in\mathbb{C}^{M}$. The received baseband signal $\mathbf{y}\in\mathbb{C}^{M}$ across the whole array is:
\begin{align}\label{eq:general_model}
\mathbf{y}=\mathbf{H}\mathbf{x}+\mathbf{n}.\vspace{-1em}
\end{align}
The BS is made of a set of $B$ sub-arrays, each of them controlled by an LPU and having $M_b=\frac{M}{B}$ antennas.
 We denote $\Tilde{\mathbf{H}}_b\in\mathbb{C}^{M_b\times K}$ and $\mathbf{y}_b\in\mathbb{C}^{M_b}$ as the channel matrix and the received signal in the $b$th sub-array for $b\in\{1,\cdots,B\}$, respectively.

\subsection{Channel Model}
We adopt the channel model presented in \cite{amiri2019message}, where the effect of VRs have been applied to the \textit{one-ring} model \cite{one_ring}. The channel for each user is
 $\mathbf{h}_k \sim \mathcal{CN}(0,\mathbf{R}_k) $ that models the small-scale fading of the channel, with channel covariance matrix of $\mathbf{R}_k$. Hence, an updated version of  $\mathbf{R}_k$ is defined as $
    [\mathbf{R}_k]_{p,q}=\frac{1}{2\Delta}\int_{-\Delta}^{\Delta}\exp{\bigl(j\mathbf{f}^T(\alpha+\theta)(\mathbf{u}_p-\mathbf{u}_q)\bigr)}\text{d}\alpha$
  for the correlation between the channel coefficients of antennas $p$ and $q$. Here, $\mathbf{f}(\omega)=-\frac{2\pi}{\lambda}\left[\cos(\omega),\sin(\omega)\right]^T$ is the wave vector with carrier wavelength of $\lambda$ and $\mathbf{u}_p,\mathbf{u}_q \in \mathbb{R}^2$ are the position vectors of the antennas $p,q$ within the VR of user $k$, angle of arrival of $\omega$ and   $\Delta$ is angular spread.
 Angle $\theta$ is the azimuth angle of user $k$ with respect to the antenna array. We have $[\mathbf{R}_k]_{p,q}=0$ when either of the antenna indices $p,q$ is outside the VR for user $k$. We use a uniform distribution for the center of the VRs within the array and the length of the VRs reads a lognormal$(\mu_l,\sigma^2_l)$ distribution \cite{gao2013massive}.

\section{Proposed Receiver Algorithms}\label{sec : BP-MF}

In this section, we describe the unified message-passing algorithm that combines the BP and mean-field approximation (MF) approaches \cite{riegler2012merging} and its application to derive our proposed decentralized receiver algorithm.%\textcolor{blue}{The MF method approximates a certain pdf by some fully factorized "simpler" pdf while the BP tries to find the marginals of a certain pdf.}
\subsection{Preliminaries}
Let $p(\bm{z})$ denote the pdf of a random vector $\bm{z}\triangleq (z_i|i\in \mathcal{I})^T$ with the set $\mathcal{I}$ indexing all the random variables in its entries. The combined BP-MF inference method can be used to calculate approximate marginals $q_i(z_i)$ that are commonly called \textit{beliefs}. To apply the BP-MF framework, first a factorization of $p(\bm{z})$ of the form
\begin{align}\label{eq: main BP-MF factorization}
p(\bm{z})=%\prod_{a\in \mathcal{A}}f_a(\bm{z}_a)=
\prod_{a\in \mathcal{A}_{\text{MF}}}f_a(\bm{z}_a)
\prod_{c\in \mathcal{A}_{\text{BP}}}f_c(\bm{z}_c),
\end{align}
should be selected, in which the different factors are classified as either belonging to the BP part (factors $f_a$, $a\in\mathcal{A}_\text{BP}$) or the MF part (factors $f_c$, $c\in\mathcal{A}_\text{MF}$). The vectors $\bm{z}_a$ denote vectors containing all the random variables that are argument of a given factor $f_a$. In addition, the sets $\mathcal{N}(a)\subseteq \mathcal{I}$ and $\mathcal{N}(i)\subseteq \mathcal{A}$ are defined to be the set of indices of all variables that are arguments of factors $f_a$ and all factors that have variable $z_i$ as an argument, respectively.

With the above factorization, an  algorithm is formulated that exchanges information, called \textit{messages}, between factors in \eqref{eq: main BP-MF factorization} and the variables $z_i, i\in \mathcal{I}$ in an iterative fashion. The messages are computed according to
\begin{subequations}\label{eq:MP equations}
\begin{align}\label{eq:MP equations1}
    m^{\text{BP}}_{a \rightarrow i}({z}_i) &=\!
    \! \int\hspace{-0.7em} \prod_{j\in\mathcal{N}(i)\setminus i}\hspace{-1em} \text{d}z_j n_{j\rightarrow a}(z_j)f_a(\bm{z}_a), \forall a\in \mathcal{A}_{\text{BP}},\\\label{eq:MP equations2}
    m^{\text{MF}}_{a \rightarrow i}({z}_i) &=\!
    \exp\bigg(\!\! \int\hspace{-0.7em} \prod_{j\in\mathcal{N}(i)\setminus i}\hspace{-1em} \text{d}z_j n_{j\rightarrow a}(z_j)\ln f_a(\bm{z}_a)\!\!\bigg)\!, \forall a\in \mathcal{A}_{\text{MF}},\\[-7pt]\label{eq:MP equations3}
    n_{i\rightarrow a}(z_i)&\propto\hspace{-2em} \prod_{c\in \mathcal{A}_{\text{BP}}\cap \mathcal{N}(i)\setminus a }\hspace{-1em}m^{\text{BP}}_{c \rightarrow i}({z}_i)\hspace{-1em}
    \prod_{c\in \mathcal{A}_{\text{MF}}\cap \mathcal{N}(i) }\hspace{-1em}m^{\text{MF}}_{c \rightarrow i}({z}_i),\forall a\in \mathcal{A}
\end{align}
 \end{subequations}
 for all $i\in\mathcal{N}(a)$, and the messages in \eqref{eq:MP equations3} are normalized so that they integrate to unity, resembling a valid pdf. From inspecting \eqref{eq:MP equations1}, we see that messages to a factor in the BP part are computed using the sum-product (SP) algorithm, while the messages to an MF factor in \eqref{eq:MP equations2} are computed using the VMP rule \cite{riegler2012merging}. At any point during the message-passing algorithms, the beliefs on each of the variables in the system can be recovered as
\begin{align}\label{eq: beliefs}
    q_i(z_i)\propto \prod_{c\in \mathcal{A}_{\text{BP}}\cap \mathcal{N}(i) }m^{\text{BP}}_{c \rightarrow i}({z}_i)
    \prod_{c\in \mathcal{A}_{\text{MF}}\cap \mathcal{N}(i) }m^{\text{MF}}_{c \rightarrow i}({z}_i),
\end{align}
 where, the beliefs are normalized to behave as proper pdfs.

\subsection{Probabilistic System Description}

\begin{figure}
	\centering
	\includegraphics[width=0.85\linewidth]{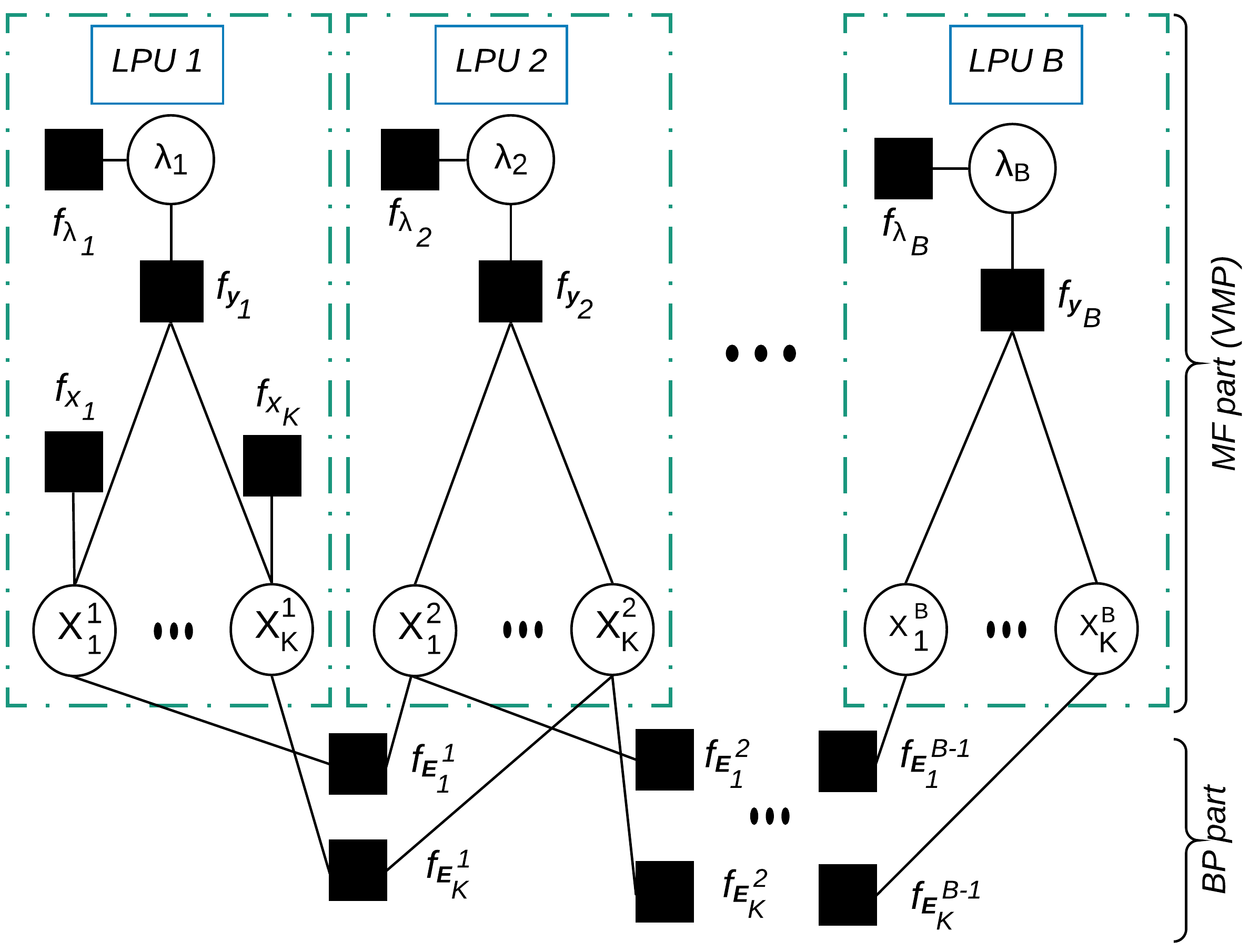}
	\caption{ \small Factor graph representation of the local processing units. Local estimations are being exchanged between the sub-arrays. 
	}
	\label{fig:ex2}
\end{figure}

We seek to detect the transmitted user symbols ${x}_k$, $k=1,\dots,K$ and estimate the noise precision (inverse of the noise variance) $\lambda=\frac{1}{\sigma_n^2}$ as a nuisance variable.
Our focus is to perform such processing locally at each of the BS sub-arrays. 
\begin{comment}

Therefore, we formulate  $B$ similar models, one for each of the sub-arrays:
\begin{align}\label{eq:joint_prob}
    p(\mathbf{x}^b,\lambda_b|\mathbf{y}_b)\propto \underbrace{p(\mathbf{y}_b|x_1^b,\cdots,x_K^b,\lambda_b)}_{f_{\mathbf{y}_b}}\underbrace{p(\lambda_b)}_{f_{\lambda_b}}\prod_{k=1}^K\underbrace{p(x_k^b)}_{f_{x_k^b}},
\end{align}\vspace{-0.5cm}
where the variables $\lambda_b$ and $\mathbf{x}^b=[x_1^b,\dots,x_K^b]^T$   represent and noise precision and the transmitted symbols observed by the $b$th BS sub-array, respectively. Also, $f_{\mathbf{y}_b}(\mathbf{x}^b, \lambda_b)=\text{CN}(\mathbf{y}_b; {\tilde{\mathbf{H}}_b\mathbf{x}^b}, \frac{1}{\lambda_b}\mathbf{I}_{M_b})$, $f_{x_k^b}(x_k^b)=\frac{1}{|\mathcal{M}|}\bm{1}_{|\mathcal{M}|}(x_k^b\in\mathcal{M})$ and $f_{\lambda_b}\propto 1/\lambda_b$ \footnote{This choice of prior corresponds to an improper, non-informative Gamma prior distribution with shape and rate parameters approaching zero.}.
\end{comment}
The posterior probability density of these two variables can be factorized as\vspace{-1em}
\begin{align}\nonumber
    p(\mathbf{x}^1&\!,\cdots\!,\mathbf{x}^B\!,\bm{\lambda}|\mathbf{y}_1,\cdots,\!\mathbf{y}_B)\!\propto\!
    p(\mathbf{y}_1|\mathbf{x}^1\!,\lambda_1)p(\lambda_1)\!\!\prod_{k=1}^K\!
    [\underbrace{p(x_k^1)}_{f_{x_k}}]\\[-15pt]
    &\times\prod_{b=2}^{B}\big[\underbrace{p(\mathbf{y}_b|\mathbf{x}^b,\lambda_b)}_{f_{\mathbf{y}_b}}\underbrace{p(\lambda_b)}_{f_{\lambda_b}}\prod_{k=1}^K\! \underbrace{p(x_k^b|x_k^{b-1})}_{f_{\text{E}_{k}^{b}}}\big]
    \label{eq:joint_prob}
\end{align}
where the variables $\lambda_b$ and $\mathbf{x}^b=[x_1^b,\dots,x_K^b]^T$   represent and noise precision and the transmitted symbols observed by the $b$th BS sub-array, respectively.  Also, $f_{\mathbf{y}_b}(\mathbf{x}^b, \lambda_b)=\text{CN}(\mathbf{y}_b; {\tilde{\mathbf{H}}_b\mathbf{x}^b}, \frac{1}{\lambda_b}\mathbf{I}_{M_b})$ and $f_{\lambda_b}\propto 1/\lambda_b$ \footnote{This choice of prior corresponds to an improper, non-informative Gamma prior distribution with shape and rate parameters approaching zero.}. $f_{x_k}=p(x_k)=\frac{1}{|\mathcal{M}|}\bm{1}_{|\mathcal{M}|}(x_k\in\mathcal{M})$ and $p(x_k^b|x_k^{b\pm1})$s are equality constraints that we will describe in \ref{section: BP section}. {Note that \eqref{eq:joint_prob} is a direct application of \eqref{eq: main BP-MF factorization} where $\prod_{k=1}^K\! p(x_k^b|x_k^{b-1})$ shows the factors for the BP part and the rest of the terms are the MF factors.}
Clearly, $\mathbf{x}^b$ and $\lambda_b$ model the same random variables for the different sub-arrays $b=1,\dots,B$. However, we treat them separately such that each sub-array can get independent estimates of them. We impose the equality constraints on the symbols $x_k$ at different sub-arrays since it is important that different sub-arrays converge to common detected symbols by exchanging messages on their respective local estimates. On the other hand, we do not use the equality constraint for the noise precision since it is a nuisance variable and, consequently, it is not critical for the algorithm if different sub-arrays yield slightly different estimates for it.
Fig.~\ref{fig:ex2} illustrates the factor graph representation of the model in \eqref{eq:joint_prob}.

\subsection{MF at the Local Processing Units}

At each of the LPUs, the processing is done using the VMP algorithm which is a message-passing interpretation of MF inference. Thus,
 the $b$th LPU aims at approximating the posterior in the distribution of $x^b$ and $\lambda_b$
 by using the approximate distribution
$
    q_b(\mathbf{x}^b,\lambda_b) = q_{\lambda_b}(\lambda_b)\prod_{k=1}^Kq_{x_k^b}(x_k^b),
$
where the na\"\i ve MF approximation is applied. To begin with, using \eqref{eq:MP equations2}, the message from factor node $f_{\mathbf{y}_b}$ to the variable node $\lambda_b$ is calculated as \cite{amiri2020distributed}
\begin{align}\label{eq:VMP_basic}
   m^{\text{MF}}_{ f_{\mathbf{y}_b\longrightarrow {\lambda_b}}}({\lambda_b})\propto {\lambda_b}^{M_b} \exp (-{\lambda_b} Z_b)
\end{align}
where  $Z_b={||\mathbf{y}_b-\sum_k \Tilde{\mathbf{h}}_{b,k} \bar{x}_k^b||^2+ \sum_k \sigma^2_{x_k^b}\Tilde{\mathbf{h}}_{b,k}^H \Tilde{\mathbf{h}}_{b,k}}$ and $\Tilde{\mathbf{h}}_{b,k}$ denotes the $k$th column of $\Tilde{\mathbf{H}}_{b}$ while $\bar{x}_k^b=\sum_{s\in\mathcal{M}}s q_{x_k^b}(s)$ and $\sigma_{x_k^b}^2 = \sum_{s\in\mathcal{M}}|s|^2 q_{x_k^b}(s)-|\bar{x}_k^b|^2$ are the mean and variance of ${x}_k^b$ with respect to $q_{x_k^b}(x_k^b)$.
Next, the LPU calculates the approximate marginal distribution $q_{\lambda_b}(\lambda_b)$ by multiplying the messages entering the variable node $\lambda_b$ as
\begin{align}\label{eq:q_lambda}
    q_{\lambda_b}\!({\lambda_b})\!\!=\!f_{\lambda_b} ({\lambda_b})\!\times\!  m^{\text{MF}}_{ f_{\mathbf{y}_b\rightarrow {\lambda_b}}}\! ({\lambda_b})=\! {\lambda_b}^{M_b-1}e^{ -\lambda_b Z_b}
\end{align}
which corresponds to a Gamma distribution with mean $
      \bar{\lambda}_b=\frac{M_b}{Z_b}.$
Afterwards, the LPU computes the messages from factor node $f_{\mathbf{y}_b}$ to the variable nodes $x_k^b$,  yielding \cite{amiri2020distributed}
\begin{align}\label{eq: f_y_to_x}
    m^{\text{MF}}_{\!f_{\!\mathbf{y}_b\!\rightarrow x^b_k}}\!\!\!\!\!\! \propto \! \mathcal{CN}\!\bigg(\!x^b_k;\! \frac{\Tilde{\mathbf{h}}_{b,k}^H}{||\Tilde{\mathbf{h}}_{b,k}||^2}
    (\mathbf{y}_b\!-\!\!\!\!\sum_{k'\neq k}\!\bar{x}^b_{k'}\Tilde{\mathbf{h}}_{b,k'}),\! \frac{1}{\bar{\lambda}_b||\Tilde{\mathbf{h}}_{b,k}||^2}\!\!\bigg)\!.
\end{align}

To conclude the MF part, the approximate marginals of the symbols of each user at the LPU $b$ are obtained by multiplying the above messages with their local priors and incoming messages from the neighboring LPUs, resulting in
\begin{align}\label{eq: SA_qx}
    q_{x^b_k}\!(x^b_k)\!\! \propto\! m^{\text{MF}}_{f_{\mathbf{y}_b}\!\rightarrow x^b_k}(x^b_k) 
    & m^{\text{BP}}_{f_{\text{E}_{k}^b}\!\rightarrow x^b_k}(x^{b}_k) m^{\text{BP}}_{f_{\text{E}_{k}^{b-1}}\!\rightarrow x^b_k}(x^{b}_k),
\end{align}
where, $m^{\text{BP}}_{f_{\text{E}_{k}^b}\rightarrow x^b_k}(x^{b}_k)$ is the message from the LPU $b+1$ to LPU $b$. 
Also, for $b=1$ the message coming from LPU $b-1$ is replaced by $f_{x_k^b}(x_k^b)$, while for $b=B$ there is no message from LPU ${b+1}$.
Basically, the equation shows that, at each LPU, the local estimate of the symbol distribution is obtained from a combination of the LPU's observed signal and the information received from the adjacent LPUs. Since adjacent LPUs estimates include, in turn, information from their respective adjacent LPUs, this mechanism ensures that after sufficient iterations of the algorithm the different LPUs converge to consistent estimates.
We describe the details for these messages in the next subsection.

\subsection{BP between the sub-arrays}\label{section: BP section}
In this subsection, we discuss the exchange of the BP messages taking place between adjacent LPUs. The fact that each LPU exchanges messages only with their neighbors, and independently of how many sub-arrays are there in total, makes our proposed solution scalable. 
To begin with, we define the equality factor nodes as 
\begin{align}
    f_{\text{E}_{k}^{b}}(x_k^b,x_k^{b+1})=\delta(x_k^b-x_k^{b+1}),
\end{align}
where $\delta(\cdot)$ is Kronecker Delta function insuring the equality of the incoming and outgoing messages to this node. Thus, we can use this function to calculate the incoming messages to the $b$th LPU from both right and left side as:
\begin{align}\label{eq: BP : right incoming}
   m^{\text{BP}}_{f_{\text{E}_{k}^b}\rightarrow x^b_k}(x^{b}_k)&=n_{ x^{b+1}_k \rightarrow f_{\text{E}_{k}^b}}(x^{b}_k),\; 1\leq b\leq B-1 \\
   m^{\text{BP}}_{f_{\text{E}_{k}^{b-1}}\rightarrow x^b_k}(x^{b}_k)&=n_{ x^{b-1}_k \rightarrow f_{\text{E}_{k}^{b-1}}}(x^{b}_k),\; 2\leq b\leq B
   \label{eq: BP : left incoming}
\end{align}
representing the incoming messages from the right and the left side, respectively\footnote{We assume that the LPUs are located and ordered from left to right.}.
Using \eqref{eq:MP equations} for the BP messages, the outgoing messages of the LPUs are computed as:
% \begin{align}\label{eq: BP: outgoing right}
%     n_{x^b_k\rightarrow f_{\text{E}_{k}^b} }(x^b_k)&= q_{x^b_k}(x^b_k)/m^{\text{BP}}_{   f_{\text{E}_{k}^b}\rightarrow x^b_k}(x^{b}_k),\; 1\leq b\leq B-1 \\
%   n_{x^b_k\rightarrow f_{\text{E}_{k}^{b-1}} }(x^b_k)&= q_{x^b_k}(x^b_k)/m^{\text{BP}}_{   f_{\text{E}_{k}^{b-1}}\rightarrow x^b_k}(x^{b}_k),\; 2\leq b\leq B \label{eq: BP: outgoing left}
% \end{align}
\begin{align}\label{eq: BP: outgoing right}
    n_{x^b_k\rightarrow f_{\text{E}_{k}^b} }\!\!(x^b_k)\!&\propto\! q_{x^b_k}\!(x^b_k)/m^{\text{BP}}_{   f_{\text{E}_{k}^b}\rightarrow x^b_k}\!(x^{b}_k),\: 1\leq b\leq B\!-\!1 \\
   n_{\!x^b_k\rightarrow f_{\text{E}_{k}^{b-1}} }\!\!(x^b_k)&\propto q_{x^b_k}(x^b_k)/m^{\text{BP}}_{\!   f_{\text{E}_{k}^{b-1}}\!\rightarrow x^b_k}\!(x^{b}_k),\; 2\leq b\leq B \label{eq: BP: outgoing left}
\end{align}
with $q_{x^b_k}(x^b_k)$ given by \eqref{eq: SA_qx}.

\subsection{Local SIC boosting}
One of the key points in the XL-MIMO non-stationary channels is uneven user energy distribution between the sub-arrays. This phenomenon can be utilized to manage the inter-user interference; symbols of the strong users in one sub-array can be detected and other sub-arrays can use this information to cancel the interference from those users.

Here, unlike conventional SIC receivers, there is no central unit to decide which users should be detected at each SIC step. Thus, we introduce a local SIC mechanism that works at each of the LPUs. Upon each update of $q_{x_k^b}(x_k^b)$, a likelihood ratio (LR) \cite{amiri2020distributed} is compared with a predefined threshold $\Gamma_{thr}$ to find the strong users. The LR metric is defined as ${\Gamma_k}\triangleq \frac{p^{(k)}_1}{p^{(k)}_2}, \forall k,$ where $p^{(k)}_1\geq p^{(k)}_2\geq\cdots p^{(k)}_{|\mathcal{M}|}$ are sorted  symbol probabilities provided by the approximate marginals $q_{x_k}\!(x_k),\;x_k\in\mathcal{M}$ of each user $k$.
\begin{comment}

\begin{prop}
The threshold value for the LR metric for any QPSK modulation can be approximated using
\begin{align}
    \Gamma_{thr}\gtrapprox \exp(\sigma^{-2}_{x})
\end{align}
where $\sigma^{2}_{x}$ is the symbol variance.
\end{prop}
\begin{proof}
Knowing that $q_{x}(x)\sim \text{CN}(x;\bar{x},\sigma^{2}_{x})$, we can rewrite the SIC condition for all the users as
\begin{align}
    \frac{\text{CN}(a^\ast;\bar{x},\sigma^{2}_{x})}{\text{CN}(a';\bar{x},\sigma^{2}_{x})}\geq\Gamma_{thr}
\end{align}
where $a^\ast$ and $a'$ are the first and second highest probable symbols, respectively. We assume that the symbol mean value $\bar{x}$ is very close to $a^\ast$ and $a^\ast$ and $a'$ are two neighbor symbols in the constellation\footnote{For a strong user with relatively low $\sigma^{2}_{x}$ value, the highest probable symbols will have the lowest distance. }.
The inequality can be proved with a few simple multiplications further. 
\end{proof}

\end{comment}

% After detecting the strong users, the LPU propagates the detected symbols to the neighboring LPUs. This is done by fixing $q_{x^b_k}(x^b_k)=\delta(x^b_k-a^\ast)$ in 
When the LR metric exceeds the threshold, i.e. when $\Gamma_k>\Gamma_{thr}$, the belief of the corresponding symbol is set to $q_{x^b_k}(x^b_k)=\delta(x^b_k-a^\ast)$, where $a^\ast$ denotes the symbol in $\mathcal{M}$ with largest probability in the approximate marginal. Note that this restriction of the belief to a delta function corresponds to a hard decision on the symbol $x_k^b$. When this belief is propagated to neighboring sub-arrays via the messages in \eqref{eq: BP: outgoing right} and \eqref{eq: BP: outgoing left}, it leads to these sub-arrays also adopting the hard decision. Analogous messages are progressively propagated to the neighboring sub-arrays, eventually yielding the same hard symbol decision for all sub-arrays.

% Eventually, after few SP message exchanges, all of the LPUs will suppress the interference from the detected users.

\subsection{The algorithm}

\begin{algorithm}[t]
\small{
	\SetAlgoLined
	\KwResult{Symbol estimates for all  active users}
	\emph{Initialize:} 
	 $M$,  $K$, $\mathbf{y}$, $B$, $\Gamma_{thr}$, $\mathcal{M}$, VMP iterations $\mathcal{J}$, total iterations $\mathcal{T}$, local detected user sets $\mathcal{S}=\{(\mathcal{S}_1=\phi),\cdots,(\mathcal{S}_B=\phi)\}$. 
	 
\For{$j = 1$ to $\mathcal{T}$ }{	

1. Initialize local user sets as $\mathcal{K}_b\triangleq\{1,\cdots,K\}$, $\forall b.$

\For{$b = 1$ to $B$ }{

2. Extract the messages to the $b$th LPU from \eqref{eq: BP : left incoming} and \eqref{eq: BP : right incoming}. 

3. Extract the channel matrix for sub-array $b$ $\Tilde{\mathbf{H}}_b$.

4. Use \eqref{eq:MRC_init} to set the initial probabilities as $q^0(x_k)$.

\For{$i = 1$ to $\mathcal{J}$ }{

5. Extract $\bar{x}^b_k$ and $\sigma^2_{x^b_k}$ values from  $q_{x_k}^{(i-1)}(x_k)$. 

6. Calculate $\bar{\lambda}_b=\frac{M_b}{Z_b}$.

7. Calculate symbol probabilities $q^{(i)}_{x_k}(x^b_k)$ using \eqref{eq: SA_qx} for all the users $k=\{1,\cdots, K\}$.
	}

8. Set the LPU's estimate as $q_{x_k}(x^b_k)=q^{(i)}_{x_k}(x^b_k)$.

\For{$k\in \mathcal{K}_b$ }{

9. Construct the LR metric $\Gamma_k$.

\uIf{$\Gamma_k\geq\Gamma_{thr}$}
{

% 10. k is a strong user; $\mathcal{K}_b=\mathcal{K}_b\setminus k$.

10.  Detect the transmitted symbol for $k$ as $\Tilde{x}_{k}$ and $\mathcal{S}_b=\mathcal{S}_b\cup \Tilde{x}_{k}$, $\mathcal{K}_b=\mathcal{K}_b\setminus k$.

11.  Cancel the interference caused by $k$ with: $\mathbf{y}_b\leftarrow \mathbf{y}_b-\Tilde{x}_{k}\Tilde{\mathbf{h}}_{b,k}$ and remove $\Tilde{\mathbf{h}}_{b,k}$ from $\Tilde{\mathbf{H}}_b$.

12. Fix the prior for $k$ as $q_{x^b_{k}}({x^b_{k}})=\delta(x^b_k-\Tilde{x}_k)$.
}
\uElseIf{$j=\mathcal{T}$ and $\Gamma_k<\Gamma_{thr}$}{

13. The user $k$ should be detected without SIC; $\Tilde{x}_{k}=\arg\max_{k\in\mathcal{K}_b} q_{x^b_{k}}({x^b_{k}})$ and $\mathcal{S}_b=\mathcal{S}_b\cup \Tilde{x}_{k}$.

14. Update the set $\mathcal{K}_b=\mathcal{K}_b\setminus k$.

}

}

15. Compute the right and left outgoing messages in \eqref{eq: BP: outgoing right} and \eqref{eq: BP: outgoing left}, respectively.

	}}

16. Choose any of the LPUs to get the detected symbol set.

	\caption{\small Proposed combined MF-BP receiver.}
	\label{alg1}}
\end{algorithm}

The proposed combined MF-BP receiver mechanism is demonstrated in Algorithm ~\ref{alg1}. 
It is composed of three main elements:
\begin{itemize}
    \item {\bf Local symbol estimation}, which is done by the VMP at each LPU (steps 3-8).
    \item {\bf The SIC detector}, that is activated if the LR metric is satisfied (steps 9-14).
    \item {\bf  Exchange of local estimates}, which takes care of the message exchanges between the LPUs using the BP (steps 2 and 15).
\end{itemize}

We use a maximal ratio combining (MRC) initialization technique \cite{amiri2020distributed} for the local estimates. Therefore, the initial approximate marginal of the symbol $x_k$ in LPU $b$ is
\begin{align}\label{eq:MRC_init}
    {q_{x_k^b}^{0}(x_k^b)} \!\propto\! {\text{CN}}\big(x_k^b; \hat x_k^b,\frac{\sum_{k'\neq k}^K \!P_{x_{k'}}
    \frac{|\Tilde{\mathbf{h}}_{b,k}^H \Tilde{\mathbf{h}}_{b,k'}|^2}{||\Tilde{\mathbf{h}}_{b,k}||^2}\! +\!\sigma_n^2}
    {||\Tilde{\mathbf{h}}_{b,k}||^2}\big)\vspace{-0.25cm}
\end{align}
which is restricted to the symbol alphabet $\mathcal{M}$. Here,
$ \hat x_k^b=\frac{\Tilde{\mathbf{h}}_{b,k}^H}{||\Tilde{\mathbf{h}}_{b,k}||^2}\mathbf{y}_b$ and $P_{x_k}$ is the user signal power.
Also, the operations in steps 13 and 14 of Algorithm 1 make sure that all the users that do not satisfy the LR condition by the end of the algorithm be detected without SIC.

\section{Performance Evaluation}\label{sec: Simulations}

\subsection{Benchmarks} 
We have implemented four different benchmarks to compare with our proposed algorithm. We choose \textit{matched filter single-user bound} which is the case when the effect of all interfering users is ideally cancelled and the target user's signal is detected by the MRC \cite{amiri2019message}. Centralized ZF as a linear and  central VMP \cite{amiri2019message} as a non-linear method are presented. Moreover, we implemented the daisy-chain algorithm from \cite{sarajlic2019fully} as a benchmark for the distributed receivers and the hierarchical VMP (H-VMP) receiver in \cite{amiri2020distributed} as a mixed method that has both local and central processing units.

\subsection{Complexity analyses}
We have calculated the computational complexity of the VMP method in one LPU in \cite{amiri2020distributed}, which is $C^{\text{MF}}=\mathcal{J}(K(4+|\mathcal{M}|)+M_b)+3M_bK$ and $\mathcal{J}$ is the number of iterations in the VMP algorithm. The additional complexity by the BP part is $2\mathcal{M}$ multiplications for computing \eqref{eq: BP: outgoing left} and \eqref{eq: BP: outgoing right} for each user and LPU\footnote{There is only one BP message for the first and last LPUs.}. Moreover, the SIC part requires $BK$ multiplications to check the LR metric. Since activation of the SIC part is not deterministic, we have a variable complexity expression that varies between the best and worst complexity cases. The worst case happens when non of the users satisfy the LR condition and $C_{\text{tot}}=\mathcal{T}\left(BC^{\text{MF}}+K[B(2\mathcal{M}+1)-2\mathcal{M}]\right)$ with $\mathcal{T}$ denoting the number of total BP iterations. The best case is when 
all of the users are detected in one iteration of the main loop with $K/B$ users detected at each LPU resulting in $C_{\text{tot}}=BC^{\text{MF}}+K(2\mathcal{M}(B-1)+1)$. The complexities for the daisy-chain, the ZF, the centralized VMP \cite{amiri2019message} and the hierarchical VMP \cite{amiri2020distributed} are
$C_{\text{DC}}=M(K+2)$, $C_{\text{ZF}}=\frac{K^3}{3}+MK^2+MK$,
$C_{\text{VMP}}=\mathcal{J}(M(3+2K)+MK|\mathcal{M}|)+3MK$ and $    C_{\text{{SIC-VMP}}}= \frac{K^2}{2}(3M+B(|\mathcal{M}|+4)+1)+MK$  , respectively. With a simple comparison, we can see that the complexity grows with a slower slope with respect to both $K$ and $M$ in $C_{\text{tot}}$ than the central methods. However, it is more complex than the daisy-chain method which is a trade-off to get a better performance and lower processing delay.

\subsection{Simulation results}
In this subsection we present the numerical results evaluating the performance of our proposed method, as well as the benchmarks. Simulation parameters are as follows: QPSK modulation, uniform linear array (ULA) with $M=300$, $K=40$, $B=5$, $\Delta=\frac{\pi}{10}$, $\theta\sim \mathcal{U}(\frac{-\pi}{2},\frac{\pi}{2})$, $\mathcal{J}=2$, $\mathcal{T}=10$, $\Gamma_{thr}=10^3$, $8\times10^4$ channel realizations and correlation matrices are updated every $50$ realizations. Centers of the VRs are uniformly distributed across the array and $\mu_l=0.7$ and $\sigma^2_l=0.2$.
 
\begin{figure}
     \centering
     \includegraphics[width=0.95\linewidth,trim={1.4cm 0 1.6cm 0.6cm },clip]{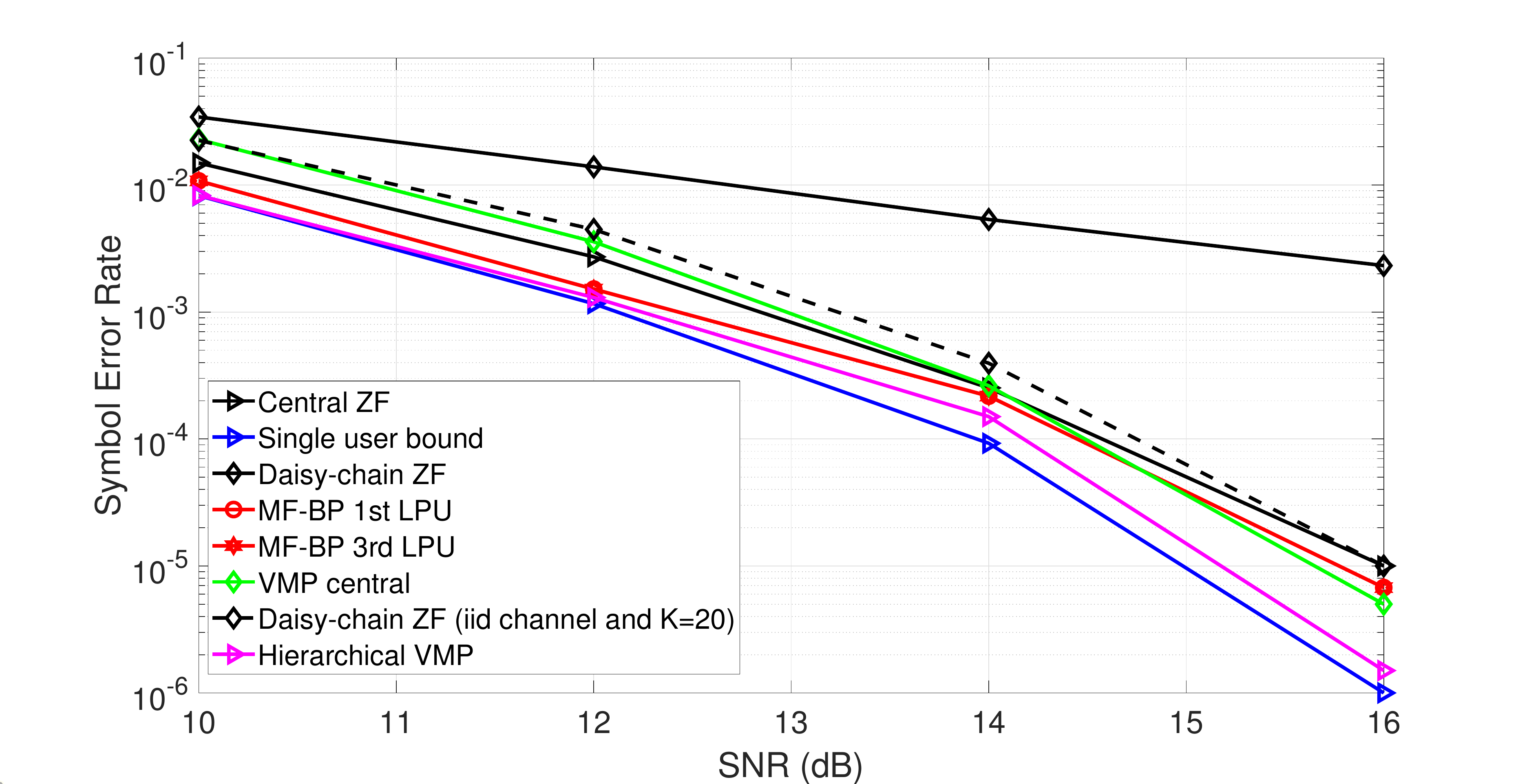}
     \caption{\small SER comparison of the proposed MF-BP method with other benchmarks. ($M=300$, $K=40$, $\Delta=\frac{\pi}{10}$) }
     \label{fig:SER_vs_SNR}
 \end{figure} 
 
 In Fig.~\ref{fig:SER_vs_SNR}, we compare the symbol error rate (SER) of the proposed method and the benchmarks. The SNR is calculated as SNR$=\frac{1}{\sigma^2_n}$ at the BS side, where we assumed unit received power for all the users over the BS array. Moreover, for our MF-BP method, we include the symbol detection results of the first and $\frac{B}{2}$th LPUs to show the convergence of the local estimates. The performance of the MF-BP method is very close to the central processing techniques and outperforms the daisy-chain technique. The reason for the degraded SER of the daisy-chain receiver is due to the channel NS and a high system load $\frac{M}{K}<10$ that makes it hard for the algorithm to completely cancel the inter-user interference. The dashed line shows the performance of the daisy-chain method in a non-correlated channel with $\frac{M}{K}=\frac{300}{20}>10$ and confirms the statement above. The H-VMP is outperforming all the methods since it uses a more complex receiver with a central SIC that repeats the detection process $K$ times.
 
 \begin{figure}
     \centering
     \includegraphics[width=0.95\linewidth,trim={1.4cm 0 1.6cm 0.6cm },clip]{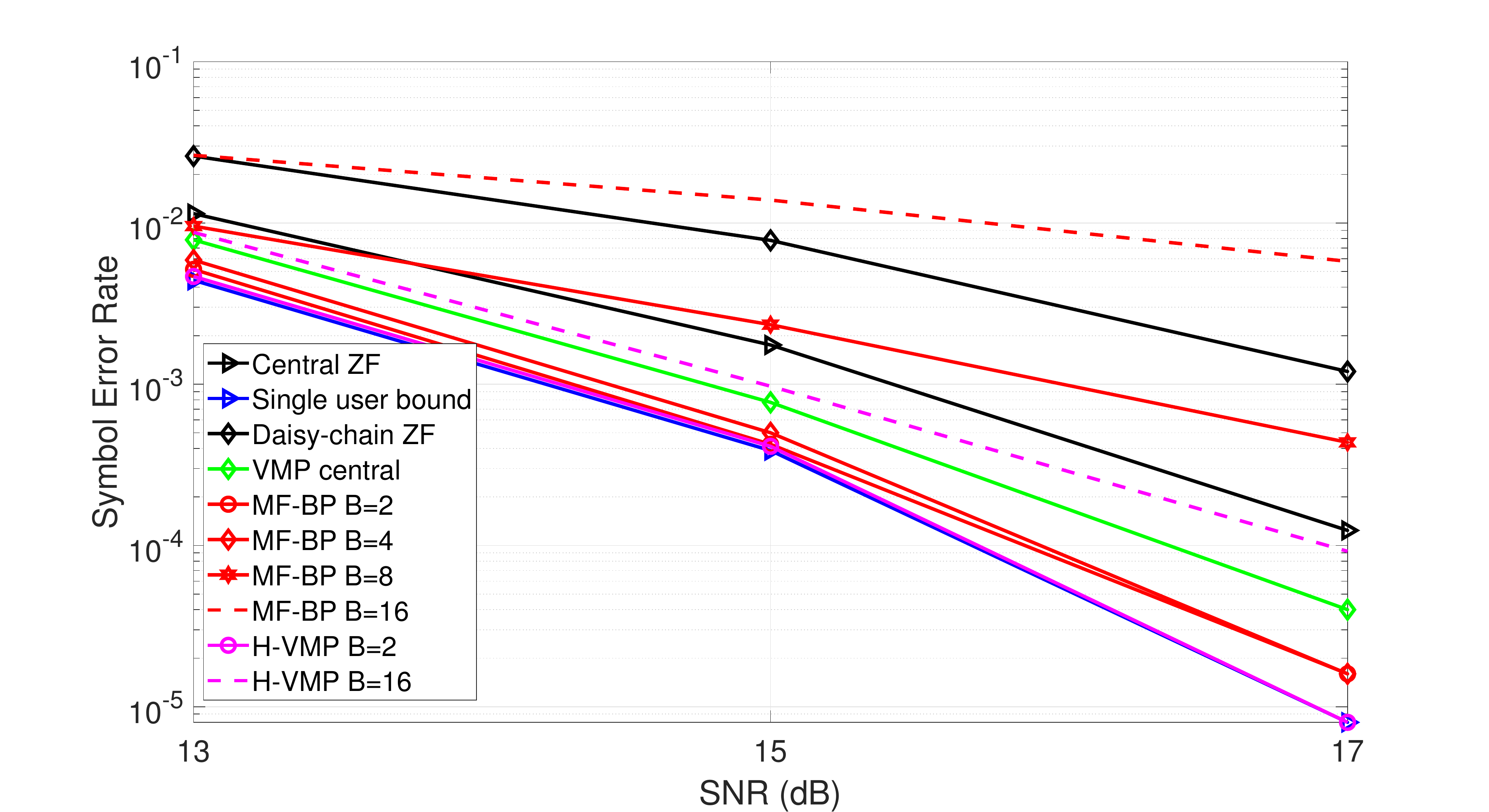}
     \caption{\small The effect of number of LPUs on the SER of the proposed MF-BP method. The curves for the single user bound and the H-VMP $B=2$ are superposed. ($M=128$, $K=25$, $\Delta=\frac{\pi}{5}$) }
     \label{fig:SER_vs_B}
 \end{figure} 
 
 The effect of number of distributed units, i.e. LPUs, on the SER of the MF-BP scheme is shown in Fig.~\ref{fig:SER_vs_B}. The performance of our method is better than the central VMP algorithm for $B=2,4$. Also, for these two values of $B$, the algorithm is giving similar results even though there are less antennas per each LPU in the latter case, i.e. $M_b|_{b=4}=\frac{1}{2}M_b|_{b=2}$. This is the result of having more SIC detectors in $B=4$ that compensate for the lower spatial resolution compared to the $B=2$ case.
 On the other hand, as $B$ increases, the number of antennas per LPU reduces (e.g. $M_b=8$ for $B=16$) while the number of users is still high ($K=25$) and the quality of local estimates weakens and results in a poor outcome.

 \section{Conclusions}
 We introduce a fully decentralized receiver for the XL-MIMO array that only relies on the LPUs for the user symbol detection. This receiver is scalable and can be deployed easily with minimum inter-connections between the sub-arrays. The ability to operate in parallel is minimizing the processing delay experienced by other benchmark techniques. Moreover, the size of exchanged messages between the LPUs, i.e. communication overhead, is very limited and allowing the use of inexpensive backhaul links. Future works will focus on integrating channel estimation and coding within the local units.

 \bibliographystyle{IEEEtran}

\end{document}